\documentclass
[aps,preprint,onecolumn,showpacs,showkeys,10pt,a4paper,floats]{revtex4}%
\usepackage{amsfonts}
\usepackage{amsmath}
\usepackage{amssymb}
\usepackage[final,dvips]{graphicx}
\usepackage[dvips]{rotating}
\usepackage{hyperref}%
\ifx\pdfoutput\relax\let\pdfoutput=\undefined\fi
\newcount\msipdfoutput
\ifx\pdfoutput\undefined\else
\ifcase\pdfoutput\else
\ifx\paperwidth\undefined\else
\ifdim\paperheight=0pt\relax\else\pdfpageheight\paperheight\fi
\ifdim\paperwidth=0pt\relax\else\pdfpagewidth\paperwidth\fi
\fi\fi\fi
\begin{document}
\title{Low-lying quadrupole collective states of the light and medium Xenon isotopes}
\author{B. Mohammed-Azizi$^{1}$, D.E. Medjadi$^{2}$}
\email{aziziyoucef@voila.fr}
\affiliation{$^{1}$University of Bechar, Bechar, Algeria. $^{2}$Ecole Normale Superieur
Kouba, Algiers, Algeria}

\begin{abstract}
Collective low lying levels of light and medium Xenon isotopes are deduced
from the Generalized Bohr Hamiltonian (GBH). The microscopic seven functions
entering into the GBH are built from a deformed mean field of the Woods-Saxon
type. Theoretical spectra are found to be close to the ones of the
experimental data taking into account that the calculations are completely
microscopic, that is to say, without any fitting of parameters.

\end{abstract}
\endpage{ }
\volumeyear{ }
\maketitle

\section{Introduction:}

The so-called General Bohr Hamiltonian ($GBH$) is applied to the light-medium
even-even Xenon isotopes region $112<A<126$. This region lies between very
deficient neutron nuclei with half-lives of few seconds and stable nuclei
$(A=124,126)$. \newline The lightest isotopes $(A=112-116)$ are not very far
away from the drip line (exotic nuclei) and hence experimental data such as
band sequences and probability transitions are difficult to obtain. For these
nuclei, we can find some data in Ref. \cite{1}-\cite{5}.\ Data of other
isotopes $(A=120-126)$ are given in Ref.\cite{6}-\cite{13}.\newline Some
simple properties can early be deduced from the concept of the equilibrium
deformation. In effect, these isotopes have a proton number $(Z=54)$ which is
somewhat close to the magic number of a closed shell $(Z=50)$ and thereby
should not contribute to an effective deformation. However, the number of
neutrons $(N=58-72)$ is far from closed shells $(N=50$ $or$ $82)$, since the
shape of the nucleus is due to both kinds of nucleons we finally should expect
an appreciable deformation for these nuclei. Experiment corroborates that
because quadrupole-beta values are found to be within the interval
$0.19<\beta<0.29$. The most deformed nuclei are the isotopes $_{{}}%
^{118-122}Xenon$ with a quadrupole deformation $\beta$ about $0.26\sim0.29$.
These nuclei correspond to the neutron mid-shell region $(N=64-66)$.\newline
An other common interesting characteristic of these nuclei is the experimental
ratio of the energy levels $R_{\exp}=E(4_{1})/E(2_{1})$ which is about
$2.3\sim2.5$ for all these isotopes. This means that these nuclei must belong
to the shape-phase transitional region between the vibrational limit $R_{\exp
}=2$ and the $\gamma-$unstable limit $R_{\exp}=2.5$.\newline There are a
number of theoretical models that attempt to explain the collective states of
the nuclei. The two main categories of these approaches are:

(1) The $GBH$ or geometrical model (see the next section) which treats the
even-even nucleus as a quantal liquid drop which vibrates and rotates with
coupling effects. It is a five dimensional Hamiltonian in which two collective
variables are devoted to the vibrations of the nucleus and three Euler angles
are used to specify its orientation with respect to the lab system. Basically
it describes the dynamic of the quadrupole deformations of the nucleus but
sometimes octupole or higher multipole orders are also considered. Some other
models such a rovibrational model (RVM) can be assimilated to a particular
case of the $GBH$ model.

(2) The interacting boson model (IBM) or algebraic model which assimilates the
even-even nucleus to some bosons outside the closed shells. These bosons are
of type d or s and are made of two nucleons strongly linked by the pairing
interaction. For odd nuclei the $IBM$ model is replaced by the interacting
bosons-fermions approximation. In fact there are two versions for this model.
In the $IBM1$, neutrons and protons are considered as the same bosons whereas
in the $IBM2$ these kind of particles are considered as distinct. $IBM$ models
can be understood only from the group theory and the Lie algebra. Symmetries
play a major role in this model. In the $IBM1$ model the vibrational,
rotational and $\gamma$-soft (or $\gamma-$unstable nuclei) limits correspond
respectively to the so-called $SU(5),U(3)$ and $SO(6)$ symmetries. Unlike the
$GBH$, the $IBM$ has a good number of particles (bosons).\newline All these
approaches contain a certain number of parameters which can be considered as
free and therefore can be fitted to experimental data to obtain the best
possible results. But these parameters can also be derived from a microscopic
theory. This is a reason why it has no sense to compare models in which
parameters are fitted to experiment data with others that are based on pure
microscopic approaches.\newline In our case the $GBH$ model contains seven
parameters or more exactly seven functions. Apart from the
macroscopic-microscopic method which is used to obtain the collective
potential energy, the other six functions are evaluated on the basis of pure
microscopic models. Similar calculations (but not exactly the same) have
already been made in the past \cite{14}-\cite{14a}.

\section{The GBH model}

Historically the Bohr Hamiltonian was established as a phenomenological model
to interpret harmonic vibrational and rotational spectra of the nuclei.
Nowadays, the term "Bohr Hamiltonian" is commonly attributed to several
similar collective hamiltonians. In this respect, we cite Ref. \cite{13a}:
"The present-day notion of the Bohr Hamiltonian is not very precise. It
encompasses a large class of Hamiltonians of which the original Bohr
Hamiltonian is only a very special case \cite{14}-\cite{25}. Here, the $GBH$
means a generic second-order differential Hermitian operator in the Hilbert
space of functions of quadrupole coordinates. It is the most general
collective Hamiltonian using the quadrupole coordinates. Making some natural
assumptions \cite{13a}, it is able to treat large amplitude collective motion.
The main advantage of the $GBH$ over the $IBM$ model is coming from the fact
that it can be derived from a microscopic theory. Two methods are usually used
to this end: (1) The adiabatic time dependent Hartree-Fock-Bogoliubov method
($ATDHFB$) which leads to quantize with some ambiguities a classical
Hamiltonian and (2) The Gaussian overlap approximation method associated with
the generator coordinates method ($GOA+GCM$) which gives straightforwardly the
quantum collective Hamiltonian. Both methods can be applied to different
microscopic models such as the mean fields models based on the Nilsson,
Woods-Saxon potentials or even self-consistent calculations with Skyrme or
Gogny effective interaction. \newline In this work we deal with the
Generalized Bohr Hamiltonian defined as a sum of three operators:%

\[
H_{\operatorname{col}}^{{}}=T_{vib}(\beta,\gamma)+T_{rot}(\beta,\gamma
,\theta_{1},\theta_{2},\theta_{3})+U_{\operatorname{col}}(\beta,\gamma)
\]
\newline where the kinetic vibrational energy and the kinetic rotational
energy are given by \cite{13a}:\newline$T_{vib}(\beta,\gamma)=-\dfrac
{\hbar^{2}}{2\sqrt{wr}}{\Huge \{}\dfrac{1}{\beta^{4}}\left[  \dfrac{\partial
}{\partial\beta}\left(  \beta^{4}\sqrt{\dfrac{r}{w}}B_{\gamma\gamma}%
\dfrac{\partial}{\partial\beta}\right)  -\dfrac{\partial}{\partial\beta
}\left(  \beta^{3}\sqrt{\dfrac{r}{w}}B_{\beta\gamma}\dfrac{\partial}%
{\partial\gamma}\right)  \right]  $

$\ \ \ \ \ \ \ \ \ \ \ \ \ \ \ \ \ \ \ \ \ \ \ \ \ \ \ \ \ \ \ \ \ \ \ \ \ \ \ \ \ +\dfrac
{1}{\beta\sin(3\gamma)}\left[  -\dfrac{\partial}{\partial\gamma}\left(
\sqrt{\dfrac{r}{w}}\sin(3\gamma)B_{\beta\gamma}\dfrac{\partial}{\partial\beta
}\right)  +\dfrac{1}{\beta}\dfrac{\partial}{\partial\gamma}\left(
\sqrt{\dfrac{r}{w}}\sin(3\gamma)B_{\beta\beta}\right)  \dfrac{\partial
}{\partial\gamma}\right]  {\Huge \}}$\newline$T_{rot}(\beta,\gamma
,\Omega)=\frac{1}{2}%
{\textstyle\sum_{k=1,2,3}}
\dfrac{I_{k}^{2}(\theta_{1},\theta_{2},\theta_{3})}{\Im_{k}}$\newline Whereas
the collective potential energy $U_{\operatorname{col}}$ of the nucleus is
defined as the potential energy of deformation of the nucleus (see the
following section):\newline$r$ and $w$ are given by: $w=B_{\beta\beta
}B_{\gamma\gamma}-B_{\beta\gamma}^{2},\;r=\Im_{1}\Im_{2}\Im_{3}$
\ \ \ \ \ \ \ \ \ \ \ \ \ \ \ \ \newline As already mentioned the $GBH$ is a
five dimentional Hamiltonian and hence contains five collective variables
$(\beta,\gamma,\theta_{1},\theta_{2},\theta_{3})$. The $GBH$ includes seven
functions: The collective energy of deformation $U_{\operatorname{col}}%
(\beta,\gamma)$, the three mass parameters $B_{\beta\beta},B_{\beta\gamma
},B_{\gamma\gamma}$ and the three moments of inertia $\Im_{1},\Im_{2},\Im_{3}$
with respect to the principal axes. All these functions are deformation
dependent. \newline The eigenvalues problem of the $GBH$ has usual the form:%

\begin{equation}
H_{\operatorname{col}}^{{}}\Psi_{\operatorname{col}}=E_{\operatorname{col}%
}\Psi_{\operatorname{col}} \label{eqbohr}%
\end{equation}

Analytical solutions of the Bohr Hamiltonian can be found in some remarkable
cases of potentials: (1) the gamma unstable nuclei, (2) the harmonic
oscillator potential, (3) the symmetric rotor model. These cases correspond
respectively to the three symmetries $SO(6)$, $U(5)$, and $SU(3)$ of the $IBM$
approach. Obviously these extremes situations are ideal cases not met in
realistic situations. Shape-phase transitional nuclei occur then between
regions of these three limiting cases. It is well know that the $GBH$ works
better for regions that are far away from closed shells. Because in the
general case there is no analytical solution, one needs then to solve
numerically the Bohr Hamiltonian. Among the numerous methods we have chose the
one of Libert (with its FORTRAN code) Ref. \cite{22}. The calculations are
done in two steps: (1) One builds the representative matrix of the Bohr
Hamiltonian with the help of a suitable basis and then (2) One diagonalizes
this Matrix. Beside the collective states obtained with this method it is also
possible to deduce other observables such as electric or magnetic transitions
probabilities, equilibrium shapes for ground state, etc...

\section{The microscopic model, some numerical details:}

In this work, single-particles energies and wave functions are obtained by the
diagonalization of the Schr\"{o}dinger equation of the stationary states.%

\begin{equation}
h\phi_{i}(\mathbf{r})=\epsilon_{i}\phi_{i}(\mathbf{r}) \label{sch}%
\end{equation}
$\epsilon_{i}$ and $\phi_{i}(\mathbf{r})$, are respectively, the eigenenergies
and the eigenfunctions of the single-particle Hamiltonian $h$%
\begin{equation}
h=-\left(  \hbar^{2}/2m\right)  \mathbf{\nabla}M_{eff}(\mathbf{r}%
)\mathbf{\nabla}+V(\mathbf{r})-\left(  \kappa/\hslash\right)  \left(
\mathbf{\nabla}W_{so}(\mathbf{r})\times\mathbf{p}\right)  \mathbf{\sigma
}+e\Phi^{Coul}(\mathbf{r}) \label{sph}%
\end{equation}
This Hamiltonian contains four contributions. They are respectively: (1) the
Kinetic energy operator, (2) the deformed central mean field, (3) the
spin-orbit contribution, and (4) the Coulomb energy for the protons. The
quantities $M_{eff}(\mathbf{r}),V(\mathbf{r}),W_{so}(\mathbf{r}),\mathbf{\Phi
}^{Coul}(\mathbf{r})$ are respectively the effective mass field, the deformed
central mean field, the deformed spin-orbit mean field and the Coulomb mean
field ($\mathbf{\sigma}$ denotes here the Pauli spin-matrices). For simplicity
we have took $M(\mathbf{r})=1.$\newline Single-particle \ hamiltonian given by
(\ref{sph}) has exactly the same structure as that obtained from the
Hartree-Fock self-consistent method with an effective nucleon-nucleon
interaction of the Skyrme III type. In our method we have simply replaced the
self-consistent one body potentials $V(\mathbf{r}),W(\mathbf{r})$ by
phenomenological deformed mean fields of the Woods-Saxon type. However for the
protons the Coulomb potential has been approximated by the one of a continuous
liquide drop model with a sharp nucleus surface. For the Woods--Saxon
potential the universal parameters' of Ref. \cite{24} have been used (see also
Ref. \cite{25}). This set of parameters claims to be able to reproduce the
correct sequences of the single-particle levels and also the nuclear
equilibrium deformations throughout the entire chart of nuclei [\cite{26},
\cite{27}]. This set is the one of $_{54}^{118}Xe_{64}$, it is given in Table
\ref{one}. \begin{table}[ptb]%
\begin{tabular}
[c]{ccc}%
neutrons &  & protons\\\hline\hline
$V_{0}=-45.99MeV$ & \multicolumn{1}{|c}{potential depth} &
\multicolumn{1}{|c}{$V_{0}=-53.22$ $MeV$}\\
$a_{V}=0.70$ $fm$ & \multicolumn{1}{|c}{potential diffuseness} &
\multicolumn{1}{|c}{$a_{V}=0.70$ $fm$}\\
$R_{V}=6.61$ $fm$ & \multicolumn{1}{|c}{potential radius} &
\multicolumn{1}{|c}{$R_{V}=6.25$ $fm$}\\\hline
$\kappa=17.74$ $MeV$ $fm^{2}$ & \multicolumn{1}{|c}{spin-orbit coupling} &
\multicolumn{1}{|c}{$\kappa=21.13$ $MeV$ $fm^{2}$}\\
$a_{so}=0.70$ $fm$ & \multicolumn{1}{|c}{spin-orbit diffuseness} &
\multicolumn{1}{|c}{$a_{so}=0.70$ $fm$}\\
$R_{so}=6.42fm$ & \multicolumn{1}{|c}{spin-orbit radius} &
\multicolumn{1}{|c}{$R_{so}=5.88$ $fm$}\\\hline
& \multicolumn{1}{|c}{charge radius} & \multicolumn{1}{|c}{$R_{ch}=6.25$ $fm$}%
\end{tabular}
\caption{{Parameters of the Woods-Saxon potential.}}%
\label{one}%
\end{table}\newline Single-particle states in Eq. (\ref{sch}) are solved by
the FORTRAN code of Ref. \cite{22a}. With these single-particle states we
calculate the potential energy surfaces of the nucleus (deformation energy) by
means of the macroscopic-microscopic method. The shell correction is
calculated by a semiclassical approach of the Strutinsky method \cite{27b}.
This means that we simply replace the Strutinsky level density by the
semiclassical one. This allows us to avoid the well-known drawbacks (smoothing
parameters) of this method. The pairing interaction is taken into account by
the $BCS$ approximation.\newline We give below some "technical" details of the
calculations.\newline The deformation energy is defined as the liquid drop
energy plus shell and pairing corrections:%

\begin{equation}
E_{def}(N,Z,\beta,\gamma)=E_{LD}(N,Z,\beta,\gamma)+\delta E_{sc}%
(N,\beta,\gamma)+\delta E_{sc}(Z,\beta,\gamma)+\delta P(N,\beta,\gamma)+\delta
P(Z,\beta,\gamma)
\end{equation}
Where the liquid drop energy is given by:%

\begin{equation}
E_{LD}(N,Z,\beta,\gamma)=\frac{3}{5}\frac{e^{2}Z^{2}}{r_{0}A^{1/3}}\left[
\frac{A}{2Z^{2}}\zeta\left[  B_{s}(\beta,\gamma)-1\right]  +\left[
B_{c}(\beta,\gamma)-1\right]  \right]
\end{equation}
in which where $B_{s}$, and $B_{c}$ are the (normalized) surface and Coulomb
contributions to the liquid drop. In the liquid drop model we have taken as in
Ref. \cite{28}: $r_{0}\approx1.275$ $fm$, and $\zeta=52.8(1-2.84I^{2})$,
$I=(N-Z)/(N+Z)$\newline The microscopic shell corrections $\delta E_{sc}$ are
evaluated separately for neutrons and protons. They are defined as the
difference between a sharp sum and a smoothed sum (between brackets) of the
energy levels:%

\begin{align}
\delta E_{shell}(N\text{ or }Z,\beta,\gamma)  &  =2\sum\epsilon_{k}%
-2\left\langle \sum\epsilon_{k}\right\rangle \\
\text{with \ \ \ \ \ \ \ \ \ }\left\langle \sum\epsilon_{k}\right\rangle  &  =%
{\displaystyle\int\limits_{-\infty}^{\lambda_{sc}}}
\epsilon g_{sc}(\epsilon)d\epsilon
\end{align}
Here $g_{sc}$ is the semiclassical level density (which by definition does not
contain shell effects ) and $\lambda_{sc}$ is the corresponding Fermi level
which is fixed by a constraint on the particle-number. The factor $2$ is due
to the time-reversal symmetry. At last, it is worth to mention that the
deformation dependence of the shell correction is contained through the
eigenvalues $\epsilon_{k}$ which depend themselves on the nuclear quadrupole
deformation $(\beta,\gamma)$.\newline The pairing correction $\delta P$ is
evaluated (also separately for protons and neutrons) with the same method as
in Ref.\cite{29}.%

\begin{equation}
\delta P_{pairing}(N\text{ or }Z,\beta,\gamma)=P-\overline{P}%
\end{equation}
where the pairing energy $P$ and the "average or smooth" pairing energy
$\overline{P}$ are given by:%

\begin{equation}
P=\underset{k=1}{\overset{N_{P}}{\sum}}2\upsilon_{k}^{2}\epsilon_{k}%
-\frac{\Delta^{2}}{G}-\overset{N_{p}/2}{\underset{k=1}{\sum}}2\epsilon
_{k}\text{, \ \ \ \ \ \ \ \ \ \ }\overline{P}=-\frac{1}{2}g_{sc}%
(\lambda)\overline{\Delta}^{2}%
\end{equation}
The number of levels used in calculations is defined in formula V3 of
Ref.\cite{29}.\newline The microscopic moments of inertia \ and mass
parameters are calculated in the usual cranking approximation of Inglis-Belyaev:%

\begin{align}
\Im_{k}  &  =2\hbar^{2}\underset{\nu,\mu}{\sum}\left\vert \left\langle
\nu\right\vert j_{k}\left\vert \mu\right\rangle \right\vert ^{2}\frac{\left(
u_{\nu}\upsilon_{\mu}-u_{\mu}\upsilon_{\nu}\right)  ^{2}}{E_{\nu}+E_{\mu}%
},\text{ \ \ }k=1,2,3\label{9119}\\
D_{ij}  &  =2\hbar^{2}\underset{\nu,\mu}{%
{\displaystyle\sum}
}\left\langle \nu\right\vert \dfrac{\partial h}{\partial i}\left\vert
\mu\right\rangle \left\langle \mu\right\vert \dfrac{\partial h}{\partial
j}\left\vert \nu\right\rangle \dfrac{\left(  u_{\nu}\upsilon_{\mu}+u_{\mu
}\upsilon_{\nu}\right)  ^{2}}{\left(  E_{\nu}+E_{\mu}\right)  ^{3}},\text{
\ \ }i,j=\beta\text{ or }\gamma\label{478}%
\end{align}
Here $i,j$ represent the quadrupole deformation parameters $\beta$ or $\gamma
$, $j_{k}$ is the single-particle angular momentum and $h$ the single-particle
Hamiltonian,. The quantity $E_{\nu}^{{}}=\sqrt{(\epsilon_{v}-\lambda
)^{2}+\Delta^{2}}$ represents as usual the energy of the quasiparticle and
$u_{\nu}^{2}=1-\upsilon_{\nu}^{2}$ is related to the occupation probability of
the level $\nu$.

\section{Results:}

\begin{figure}[ptbh]
\includegraphics[angle=-00,width=120mm,keepaspectratio]{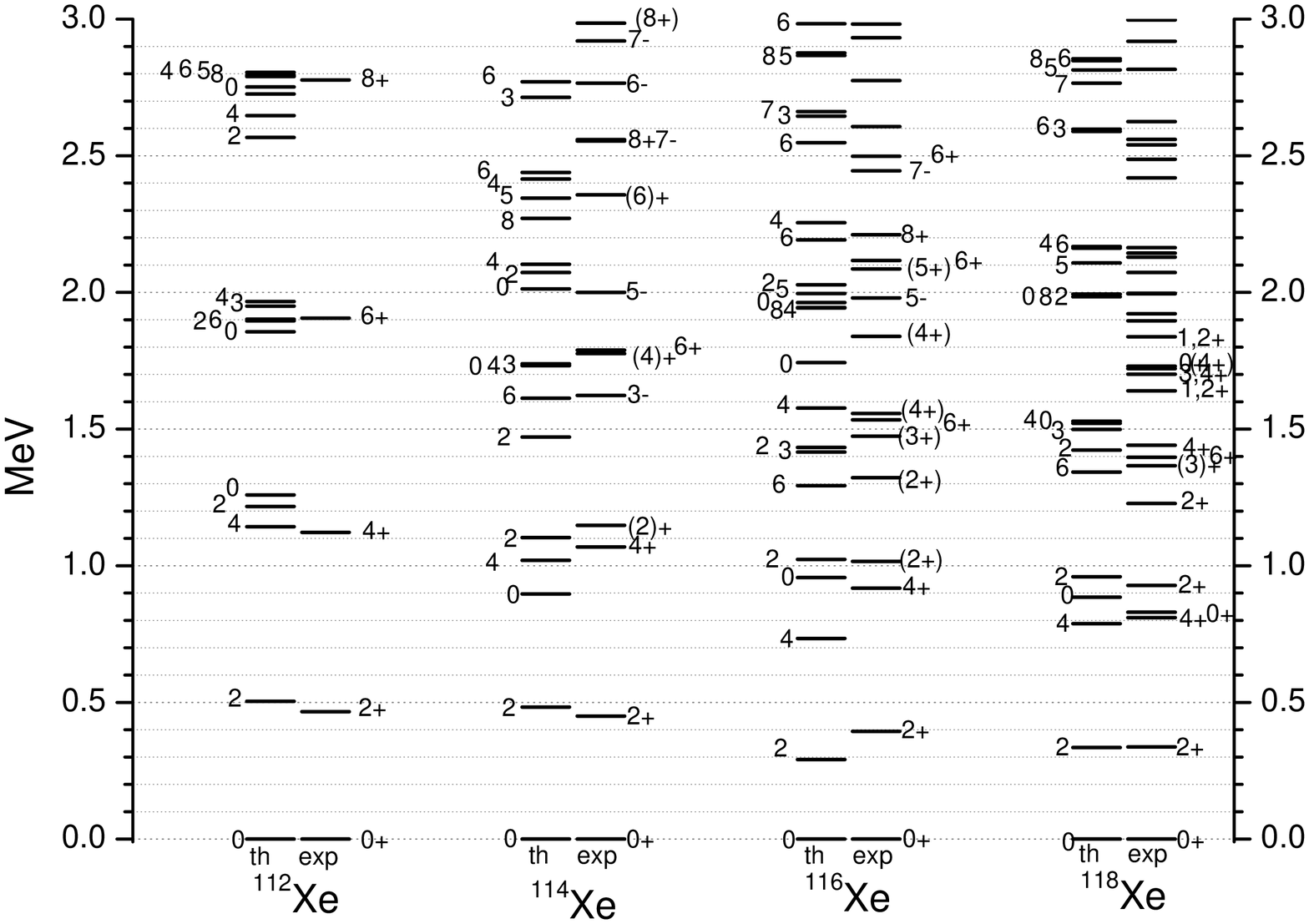}\caption{Theoretical
collective levels calculated by the General Bohr Hamiltonian with potential
energy surface and inertials functions evaluated by means of microscopic
method using the Woods-Saxon potential. We considere here only quadrupole
collective level. The parity of the theoretical levels is therefore positive.
This is the reason why it is simply omitted.}%
\label{spectrum1}%
\end{figure}

\begin{figure}[ptbhptbh]
\includegraphics[angle=-00,width=120mm,keepaspectratio]{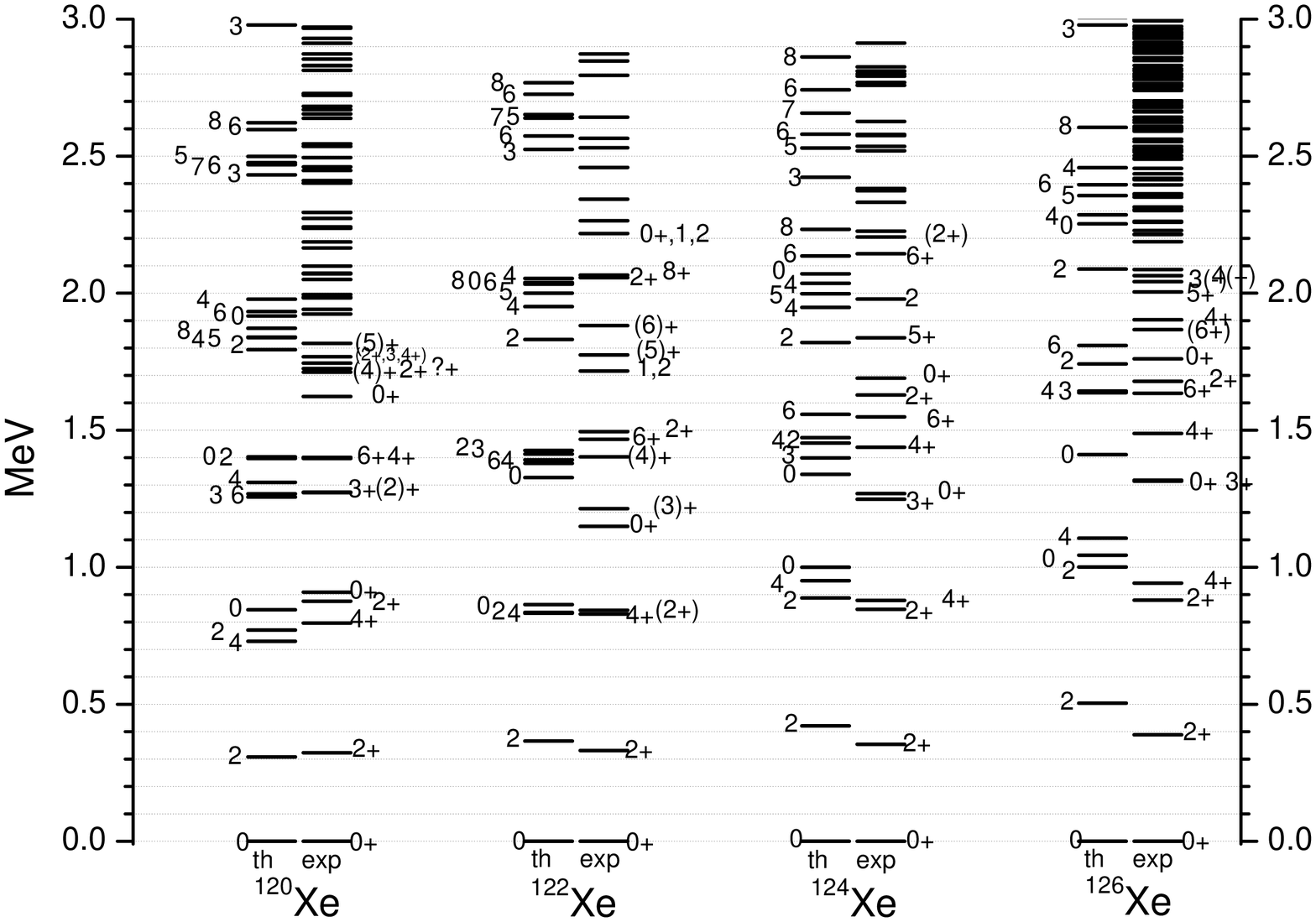}\caption{Continution
of Fig. 1}%
\label{spectrum2}%
\end{figure}

\begin{figure}[ptbh]
\includegraphics[angle=0,width=90mm,keepaspectratio]{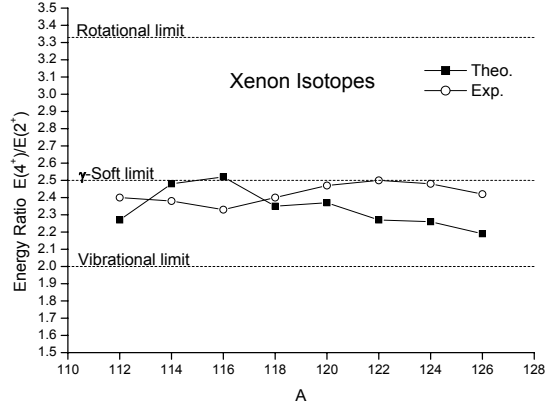}\caption{Energy
ratio $E(4_{1}^{+})/E(2_{1}^{+})$ for the family of isotopes $_{54}%
^{112-126}Xe$. }%
\label{ratio}%
\end{figure}

From numerous calculations based on microscopic evaluations of the inertial
functions (i.e. with no free parameters in the $GBH$), it turns out that
generally real difficulties are encoutered to reproduce properly the
experimental collective levels. Among these difficulties is the fact that the
scales of theoretical collective spectra are generally too stretched compared
to the experimental ones. "Fine structure" such as the order of the levels and
their relative positions become then a challenge and it is not easy to correct
these defects. This is because the seven functions of the $GBH$ cannot be
deduced directly from experiment and hence cannot be known without ambiguity.
Moreover it often happens for the same model that good predictions in
theoretical spectra are not "corroborated" by good probability transitions and
vice-versa.\newline In the following our theoretical spectra will be compared
to the ones deduced experimentally and compiled in the websites:
$http://www.nndc.bnl.gov/endsf.\newline$Our main results can be summarized as follows:

The spectra obtained by our calculations are given in figures \ref{spectrum1}
and \ref{spectrum2} at the left hand side for each isope. The theoretical low
energy levels are grouped in sets.\newline\newline They are characterized by a structure which is very close
to that of the five dimensional (anharmonic) vibrator . The triplets
$(0^{+},2^{+},4^{+})$ and the quintuplets $(0^{+},2^{+},3^{+},4^{+},6^{+})$
are as a rule, present for all the isotopes.$^{112-126}Xe$. However,
experimental data given at the right hand side for each isotope show that only
two nuclei belong to this type, namely $^{118}Xe$ and $^{120}Xe$ which are
well described here . All other nuclei are characterized by doublets
$(4^{+},2^{+})$. This suggests that these nuclei have a spectrum structure
close to the one of the $\gamma$ unstable nuclei of Wilets-Jean model which
predicts the doublet $(4^{+},2^{+})$. \newline We recall that the Wilets-Jean
model is essentially based on the hypothesis that the potential energy does
not depend on the axial assymmetry $\gamma$ \cite{32}, i.e. $\partial
U_{\operatorname{col}}(\beta,\gamma)/\partial\gamma=0$ with a strong minimum
out of the spherical shape (i.e. the minimum occurs for $\beta\neq0$). These
nuclei are oten called as to be $\gamma-$soft. In this respect, it is possible
to consider the harmonic vibrator as a particular case of the Wilets-Jean
model. In effect, the Wilets-Jean model is defined by the condition $\partial
U_{\operatorname{col}}(\beta,\gamma)/\partial\gamma=0$ which is filled by the
harmonic vibrator. The only difference comes from the fact that the minimum of
the collective potential energy occurs for the spherical shape whereas in the
W-J model it lies elsewhere. Consequently, it is then not surprising that
among them two are of vibrational type. \newline It is easy to explain why our
theoretical values are close to the ones of the anharmonic vibrator:\newline%
\qquad(i) The potential enegy depends very little on the axial assymmetry
parameter $\gamma$ (as for Wilets-Jean model)\newline\qquad(ii) The
deformation energy (defined as the difference between the energy for the
spherical shape and the one obtained for the equilibrium deformation) is too
weak. In other words these nuclei are theoretically too "soft" in the $\beta$
degree of freedom and the well is insufficiently pronounced for $\beta\neq0$
to obtain Willets-Jean potential type.\newline\qquad(ii) Moreover the mass
parameters $B_{\beta\beta},B_{\beta\gamma},B_{\gamma\gamma}$vary very little
in the vicinity of the minimum. Therefore as in the original Bohr model the
mass parameters can be considered as constant.\newline Above the doublets we
find either a set of the type $(2^{+},3^{+},4^{+},6^{+})$ or of the type
$(0^{+},2^{+},3^{+},4^{+},6^{+}).$ The experimental energy ratios $E(4_{1}%
^{+})/E(2_{1}^{+})$ lie between the the $\gamma-$soft and the vibrational
limits in Fig.\ref{ratio} but are much more closer to the $\gamma-$soft limit
especially for the last isotopes. The theoretical values are not so far away
from the experimental ones but their behaviours as a function of A seem to be
opposite from each other.

\begin{figure}[ptbh]
\includegraphics[angle=0,width=90mm,keepaspectratio]{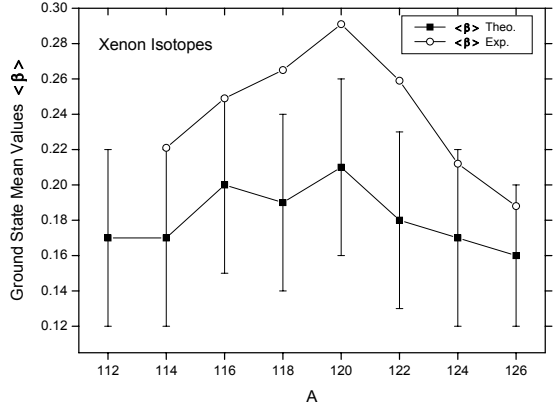}\caption{Root
mean Square value $\sqrt{<\beta^{2}>}$of the quadrupole deformation in the
ground state as function of the mass number for the Xenon isotopes. }%
\label{meanbeta}%
\end{figure}

The first satisfaction of these calculations is the correct scale of the
spectra without any kind of correction. In other words the usual defect of the
stretching \cite{13a} of the spectra is not present. This seems essentially
due to the Woods-Saxon mean field which gives the right values for the mass
parameters. In earlier similar works, for example the one of \ Ref.\cite{14}
the stretching of the collective spectra is "cured" by introducing the pairing
vibrations in the calculations whereas in Ref. \cite{14a} it is corrected by
reducing artificially the pairing strenghs by $20\%$. The second satisfaction
is the calculated levels $2_{1}^{+},$and the doublets $\left(  4_{1}^{+}%
,2_{1}^{+}\right)  $ which are for the most nuclei very close to the
experimental ones. Moreover as already noted the energy ratios given by our
model in Fig.\ref{ratio} are quite close to the experimental ones.\newline
Nevertheless we obtain also bad results such as for the mean values of the
ground states quadrupole deformation $\left\langle \beta\right\rangle $ in
Fig.\ref{meanbeta}. The experimental values are deduced from transitions
probabilities $B(E2;0^{+}\rightarrow2^{+})$ from the website
$http://www.nndc.bnl.gov/be2$ whereas the deformations given by the FORTRAN
code come from the collective wave function ($B(E2;0^{+}\rightarrow2^{+})$
values are not calculated by the present version of the code).

\section{Conclusion}

Microscopic calculations based on the Woods-Saxon mean field were performed to
find the potential energy and the six inertial functions entering into the
collective Bohr hamiltonian. We have considered the light and medium Xenon
isotopes of the region $112<A<126$. Then, the Bohr hamiltonian has been
diagonalized without any fitting of parameters. The resulting spectra were
found to be quite close to the experimental ones. Moreover the values of the
mass parameters seemed to be correct in magnitude giving a good scaling in the
collective spectra for this region. Most of theoretical collective levels
$E(2_{1}^{+})$ \ were found to be in good agreement with the experimental
ones. The theoretical energy ratios $E(4_{1}^{+})/E(2_{1}^{+})$ were also
fairly well. However, the major difference comes from the fact that our
spectra belong to the (anharmonic) vibrator type whereas most of experimental
spectra exhibit doublets of \ the Wilets-Jean model. Both alike are of the
same type ($\gamma-$soft), however contrarily to the Wilets-Jean model, our
collective potential energy does not posses a sharp minimum for a deformed
shape $(\beta\neq0)$. In other words the shell correction is not sufficiently
strong to modify significantly the potential energy of the liquid drop model.
Consequently some further studies seem to be necessary in order to correct
this defect.

\end{document}